\documentstyle[12pt]{article}
\def\PR #1 #2 #3 {Phys.~Rev.~{\bf #1}, #2 (#3)}
\def\PRL #1 #2 #3 {Phys.~Rev.~Lett.~{\bf #1}, #2 (#3)}
\def\PRD #1 #2 #3 {Phys.~Rev.~D~{\bf #1}, #2 (#3)}
\def\PLB #1 #2 #3 {Phys.~Lett.~{\bf B#1}, #2 (#3)}
\def\NPB #1 #2 #3 {Nucl.~Phys.~{\bf B#1}, #2 (#3)}
\def\RMP #1 #2 #3 {Rev.~Mod.~Phys.~{\bf #1}, #2 (#3)}
\def\ZPC #1 #2 #3 {Z.~Phys.~C~{\bf #1}, #2 (#3)}

\begin{document}
 
\rightline{hep-ph/9512292}
\rightline{ILL-(TH)-95-33}
\medskip
\rightline{December 1995}
\bigskip\bigskip

\begin{center} {\Large \bf Spin Correlation in Top-Quark Production \\
\medskip at Hadron Colliders.} \\ 
\bigskip\bigskip\bigskip\bigskip
{\large{\bf T.~Stelzer} and {\bf S.\ Willenbrock}} \\ 
\medskip Department of Physics\\
University of Illinois \\ 1110 West Green Street \\  Urbana, IL 61801\\
\bigskip 
\end{center} 
\bigskip\bigskip\bigskip

\begin{abstract}
We propose techniques to observe the correlation of the spins of top 
quarks and antiquarks at the Tevatron and the LHC.  Observation of the spin
correlation would confirm that the top quark decays before its spin 
flips, and would place a lower bound on the top-quark width and $V_{tb}$. 
The spin correlation may also be a useful tool to study the weak decay 
amplitude of the top quark.
\end{abstract}

\addtolength{\baselineskip}{9pt}

\newpage

\indent\indent 
The discovery of the top quark at the Fermilab Tevatron by the 
CDF and D0 collaborations \cite{TOP} has ushered in the era
of top-quark physics.  The large mass of the top quark, $m_t = 180 \pm 12$
GeV, in comparison with the five lighter quarks, suggests it may play a
special role in particle physics.  It is thus imperative that we examine
the physics of the top quark in detail.

Run II at the Tevatron, beginning in 1999, will provide each experiment
with approximately 1000 fully-reconstructed, $b$-tagged $t\bar t$
events \cite{Tev2000}. Even higher yields will become available from the 
CERN Large Hadron Collider (LHC) \cite{Aachen} and possible upgrades of
the Tevatron \cite{Tev2000}. These large statistics should allow a detailed 
study of the properties of the top quark.

The standard model predicts that the top quark decays before its spin 
flips \cite{BDKKZ}.  This is in contrast with the lighter quarks,
which are depolarized by QCD interactions long before they 
decay \cite{FP}.\footnote{An exception is the production of heavy baryons 
\cite{MS}.}
The spin of the top quark is therefore reflected by its decay products.

While the top quarks and antiquarks produced at hadron colliders are
unpolarized, their spins are {\em correlated} \cite{BOP}.  Figure 1
shows the cross sections for the production of $t\bar t$ pairs with
the same and opposite helicities at the Tevatron and the LHC, as a 
function of the $t\bar t$ invariant mass, for $m_t=175$ GeV.\footnote{The 
spin correlation is easier to observe the larger the top-quark mass, so we
use $m_t=175$ GeV throughout, to be conservative.} 
The helicity is the spin component along the particle's momentum, in the 
$t\bar t$ center-of-mass frame. At the Tevatron, $70\%$ of 
the pairs have the opposite helicity, while $30\%$ have the same 
helicity. Defining the correlation as\footnote{Parity conservation in QCD 
implies 
$\sigma(t_R\bar t_L)=\sigma(t_L\bar t_R)$, and CP conservation implies 
$\sigma(t_R\bar t_R)=\sigma(t_L\bar t_L)$ \cite{PS}.}
\begin{equation}
C\equiv\frac{\sigma(t_R\bar t_R+t_L\bar t_L)-\sigma(t_R\bar t_L+t_L\bar t_R)}
{\sigma(t_R\bar t_R+t_L\bar t_L)+\sigma(t_R\bar t_L+t_L\bar t_R)}
\label{correl}
\end{equation}
we find that the helicities of the top quarks and antiquarks have a 
correlation of $-40\%$.  At the LHC the correlation is $+31\%$. 
Figure 1 also shows that placing cuts on the invariant mass of the
$t \bar t$ system can enhance the correlation at both the Tevatron and 
the LHC.

In this paper we propose techniques for observing the correlation of
the top-quark and -antiquark helicities experimentally. There are two
motivations for doing so.  First, observation of the spin correlation
would confirm that the top quark does indeed decay before its spin
flips, thereby setting an upper bound on the top-quark lifetime.  This
would in turn place a lower bound on the top-quark width,\footnote{A
measurement of the invariant mass of the top-quark decay products
places an {\em upper} bound on the width.} which is proportional to
the combination of Cabbibo-Kobayashi-Maskawa (CKM) matrix elements
$|V_{td}|^2+|V_{ts}|^2+|V_{tb}|^2$.  If there are just three
generations of quarks this quantity equals unity, but it can be almost
zero if there are more than three generations.

The spin of a heavy quark is flipped by its chromomagnetic moment, which 
is inversely proportional to its mass, $m_Q$.  The spin-flip time is 
therefore proportional to $m_Q/\Lambda_{QCD}^2$.  The chromomagnetic 
moment is also responsible for the hyperfine splitting in heavy mesons, 
so this splitting can be used to estimate the spin-flip time \cite{FP,PS}.  
Scaling from the $D$--$D^*$ and $B$--$B^*$ mass splittings, we estimate the 
spin-flip time of the top-quark to be $(1.3\;{\rm MeV})^{-1}$.  This is much 
longer than the anticipated top-quark lifetime, 
$\Gamma^{-1}\approx(1.5\;{\rm GeV})^{-1}$, assuming three generations.  

Assuming more than three generations, observation of the spin
correlation would imply $|V_{td}|^2+|V_{ts}|^2+|V_{tb}|^2 > (0.03)^2$.
If we assume $|V_{tb}|$ is much larger than $|V_{td}|$ and
$|V_{ts}|$,\footnote{This assumption can be tested by measuring the
ratio $\frac{|V_{tb}|^2}{|V_{td}|^2+|V_{ts}|^2}$, which can be
extracted by comparing the number of single- and double-$b$-tagged
$t\bar t$ events \cite{Tev2000}.}  this yields $|V_{tb}| > 0.03$.  If
$|V_{tb}|$ proves to be less than this bound, it would mean that the 
recently-discovered ``top'' quark is not the SU(2) partner of the bottom quark,
and that the real top quark is still at large.

The second motivation is that we envision that the spin correlation can be
used to help probe for non-standard interactions in the weak decay of
the top quark \cite{KLY}. The weak decay does not affect the
correlation, which arises from QCD, but it does affect how the
correlation manifests itself in the top-quark decay products.
Non-standard weak interactions of the top quark could result from the
mechanism which provides the top quark with its large mass \cite{PZ}.
In this paper we restrict our attention to the spin correlation in the
standard model, since one must first establish that it is observable
in that case.

The dominant production mechanism for $t \bar t$ pairs at the Tevatron
is $q \bar q \rightarrow t \bar t$, which proceeds through a $J=1$
$s$-channel gluon.  Near threshold, the $t \bar t$ pair has zero orbital
angular momentum, so the $t\bar t$ pair is
in a $^3S_1$ state \cite{AS}, with spin eigenstates
\begin{eqnarray*}
|+ +\rangle \;\\
{1\over \sqrt{2}} \left[ |+-\rangle + |-+\rangle \right] \\
|- - \rangle \;.
\end{eqnarray*}
Since the $t$ and $\bar t$ move oppositely in the $t \bar t$
center-of-mass frame, they have the opposite helicity if they have the
same spin, and the same helicity if they have the opposite spin.  Two
of the three states have the opposite helicity, hence the correlation
near threshold is $C = +{1\over 3}-{2\over 3} = -33\%$.  Far above
threshold, helicity conservation at high energy ensures that the $t$
and $\bar t$ are produced with the opposite helicity, so $C=-100\%$.
The formula that interpolates between the two extremes is
\begin{equation}
\frac{\sigma(t_R\bar t_L+t_L\bar t_R)}{\sigma(t_R\bar t_R+t_L\bar t_L)}
= 2 \frac{M_{t\bar t}^2}{4 m_t^2} \;.
\end{equation}
Convoluting with parton distribution functions, 
and including the small contribution from $gg\to t\bar t$, yields 
the Tevatron curves in Fig.~1.  Integrating over the $t\bar t$ invariant 
mass yields an average correlation of $-40\%$.

At the LHC the situation is reversed. The dominant contribution to the
cross section comes from $gg \to t\bar t$.  Near threshold, the $t\bar
t$ pair is in a $^1S_0$ state \cite{H,AS}
\begin{eqnarray*}
{1\over \sqrt{2}} \left[ |+-\rangle - |-+\rangle \right] \;.
\end{eqnarray*}
The $t$ and $\bar t$ therefore have the same helicity, with a
correlation of $+100\%$. Far above threshold, helicity conservation
again ensures that the $t$ and $\bar t$ are produced with the opposite
helicity.  Convoluting with parton distribution functions, and
including the small contribution from $q\bar q\to t\bar t$, yields the
LHC curves in Fig.~1.  The average correlation, integrating over the
$t\bar t$ invariant mass, is $+31\%$.

Since the ratio of the same- and opposite-helicity cross sections
varies with the $t\bar t$ invariant mass, the correlation at both the
Tevatron and the LHC can be enhanced by cutting on this quantity.  A
cut of $M_{t\bar t}>415$ GeV increases the correlation to $-50\%$ at
the Tevatron, with an acceptance of $55\%$. A cut of $M_{t\bar t}<475$
GeV increases the correlation to $+50\%$ at the LHC, with an
acceptance of $45\%$.

We now consider how the top quark's helicity is reflected by 
its decay products, either $t\to b\ell^+\nu$ or $t\to bu\bar d$.  
In the rest frame of the parent top quark, the angular
distribution of fermion $i$ with respect to the momentum of the 
top quark in the $t\bar t$ center-of-mass frame is \cite{J}
\begin{equation}
\frac{dN_{R,L}}{d \cos \theta^*_i} = \frac{1}{2}(1 \pm h_i \cos \theta^*_i) 
\label{decay}
\end{equation}
where $h_i$ is a constant between $-1$ and 1.  The ability to distinguish
$t_R$ from $t_L$ evidently increases with $|h_i|$.  For top 
antiquarks, the subscripts $R,L$ are interchanged in Eq.~(\ref{decay}).

The most powerful spin
analyzer is the charged lepton in semi-leptonic decay, for which 
$h_\ell=1$ \cite{J,JK}.  Similarly, in hadronic decay, the $\bar d$ has 
$h_d=1$.  Unfortunately, it is impossible to distinguish the $u$ and 
$\bar d$ jets.  However, half of the hadronic decays are $t\to bc\bar s$, 
and one might be able to tag the charm quark. The high analyzing power of 
this decay might compensate the charm-tagging efficiency, which is 
unknown at present.

The $b$ quark has $h_b=-0.41$, which can be derived as follows.  
If the $W$ boson is longitudinal, $h_b=-1$; if it is transverse, 
$h_b=+1$.  Since $70\%$ of the $W$ bosons in top decay are longitudinal, 
the net value of $h_b$ is approximately $-0.4$ \cite{J}.

For hadronic decay, one can use the least-energetic quark (in the
top-quark rest frame) from the $W$ decay, which has $h_q=+0.51$.  This
follows from the fact that the $\bar d$ quark is the least-energetic
quark $61\%$ of the time.  Since the $\bar d$ quark has the greatest
analyzing power, the least-energetic quark has significant analyzing
power \cite{J}.

The angular distribution of fermion $i$ from the $t$ decay and fermion
$j$ from the ${\bar t}$ decay in $t\bar t$ events is\footnote{This
equation assumes the cross section factorizes into production times
decay, maintaining the spin correlation, but neglecting interference
effects.  We have checked that the interference effects are indeed
small.}
\begin{equation}
\frac{d^2N}{dz_idz_j}=(1-z_iz_jh_ih_jC)
\end{equation}
where
\begin{equation}
z_i = \cos \theta^*_i
\end{equation}
and where $C$, defined in Eq.~(\ref{correl}), is the degree of spin
correlation.  For uncorrelated events $C = 0$, and the distribution is
flat in the $z_i z_j$ plane.  A simple measure of the
correlation is the asymmetry in the $z_iz_j$ plane.  We find
\begin{equation}
A \equiv \frac{N_+-N_-}{N_++N_-} = -\frac{1}{4} h_i h_j C
\label{asymm}
\end{equation}
where $N_+$ is the number of events with the product $z_i z_j > 0$ and
$N_-$ is the number of events with $z_i z_j < 0$.  The largest
asymmetry is obtained by maximizing the product $|h_ih_j|$.  Since the
lepton in semi-leptonic decays has $h_\ell=1$, we always use one
semi-leptonic decay.

For dilepton events, evaluating Eq.~(\ref{asymm}) at the Tevatron
($C=-40\%$) yields $A=+10\%$. However, since the dilepton events have
two neutrinos, the events are not fully reconstructable, and therefore
are not amenable to our analysis.  This asymmetry represents the
theoretical upper bound.  Other methods for observing spin correlation
with dilepton events are discussed in Refs.~\cite{BOP,H,AS,CLT}.

Let us concentrate on the fully-reconstructable $W+4$ jet events,
where the $W$ boson decays leptonically.  An asymmetry of $A=+10\%$ is
achievable via charm tagging, but the efficiency of this is unknown at
present.  If the efficiency is $\epsilon$, the number of charm-tagged
events in 1000 $W+4$ jet events is $500\epsilon$.  The significance of
the asymmetry (its difference from zero) is thus
$10\%\times\sqrt{500\epsilon}=2.2\sqrt{\epsilon}\sigma$.

The next most powerful spin analyzer is the least-energetic quark.
This yields an asymmetry of $A=+5.1\%$, which has a significance of
$1.6\sigma$ with 1000 events.  Using the $b$ quark yields $A=-4.1\%$,
which has a significance of $1.3\sigma$ with the same number of
events.  We have found that these measurements are uncorrelated, so we
may combine the two to increase the significance.  This may be
understood by recalling that the $b$-quark's analyzing power arises
from longitudinal $W$ bosons, and is degraded by the transverse $W$
bosons.  On the other hand, the least-energetic quark's analyzing
power arises from both the longitudinal and the transverse $W$ bosons.
Combining the two measurements, the significance of 1000 $W+4$ jet
events is $2\sigma$.

We conclude that the top-quark spin correlation is potentially
observable, at the $2\sigma$ level, in Run II at the Tevatron.
Increasing the integrated luminosity of the Tevatron by a factor of 10
increases the significance by a factor of three, ensuring observation
of the correlation, and perhaps allowing the correlation to be used as
a tool to study the weak decay amplitude of the top quark.  At the
LHC, the top-quark cross section is about 100 times greater than at
the Tevatron.  That, combined with the anticipated large integrated
luminosity, will result in at least a million fully-reconstructed,
$b$-tagged $t\bar t$ events.  The spin correlation is potentially
measurable to one percent, and it may be a powerful tool to study the
weak decay amplitude of the top quark with high precision.

The results in this paper are based on leading-order parton-level
calculations of the signal.  Further work is needed to determine the
effect of higher-order corrections, backgrounds, hadronization, and
detector response on this analysis.

Note: A recent paper by Mahlon and Parke also studies top-quark spin
correlation at hadron colliders \cite{MP}.  Our work agrees with
theirs where there is overlap.

\section*{Acknowledgements}
\indent\indent We are grateful for conversations with
S.~Errede, R.~Gardner, T.~Liss, and J.~Thaler.  This work was
supported in part by Department of Energy grant DE-FG02-91ER40677.

\clearpage

\section*{Figure Captions}

\vrule height0pt \vspace{-22pt}

\bigskip

\indent Fig.~1 - Cross sections for $t\bar t$ with the same helicities 
$(t_R\bar t_R+t_L\bar t_L)$ and the opposite helicities $(t_R\bar
t_L+t_L\bar t_R)$ at the Tevatron ($\sqrt s=2$ TeV $p\bar p$ collider)
and the LHC ($\sqrt s=14$ TeV $pp$ collider), versus the $t\bar t$
invariant mass.  The MRS(A$^\prime$) parton distribution functions
were used \cite{MRSA}.


\begin{thebibliography}{99}

\bibitem{TOP} CDF Collaboration, F.~Abe {\it et al.}, \PRL 74 2626 1995 ;
D0 Collaboration, S.~Abachi {\it et al.}, \PRL 74 2632 1995 .

\bibitem{Tev2000} {\sl Report of the Tev2000 Study Group on Future 
Electroweak Physics at the Tevatron}, 
eds.~D.~Amidei and R.~Brock, D0 Note 2589/CDF Note 3177 (1995).

\bibitem{Aachen} {\sl Proceedings of the ECFA Large Hadron Collider Workshop},
Aachen, Germany, 1990, edited by G.~Jarlskog and D.~Rein (CERN Report 
No.~90-10, Geneva, Switzerland, 1990).

\bibitem{BDKKZ} I.~Bigi, Y.~Dokshitzer, V.~Khoze, J.~K\"uhn, and 
P.~Zerwas, \PLB 181 157 1986 .

\bibitem{FP} A.~Falk and M.~Peskin, \PRD 49 3320 1994 .

\bibitem{MS} T.~Mannel and G.~Schuler, \PLB 279 194 1992 ; F.~Close, 
J.~K\"orner, R.~Phillips, and D.~Summers, J.~Phys.~{\bf G 18}, 1716 (1992).

\bibitem{BOP} V.~Barger, J.~Ohnemus, and R.~Phillips, 
Int.~J.~Mod.~Phys.~{\bf A4}, 617 (1989).

\bibitem{MRSA} A.~Martin, R.~Roberts, and W.~J.~Stirling, \PLB 354 155 1995 .

\bibitem{PS} M.~Peskin and C.~Schmidt, \PRL 69 410 1992 .

\bibitem{KLY} G.~Kane, G.~Ladinsky, and C.-P.~Yuan, \PRD 45 124 1992 .

\bibitem{PZ} R.~Peccei and X.~Zhang, \NPB 337 269 1990 .

\bibitem{H} Y.~Hara, Prog.~Theor.~Phys.~{\bf 86}, 779 (1991).

\bibitem{AS} T.~Arens and L.~Sehgal, \PLB 302 501 1993 .
 
\bibitem{J} M.~Je\.zabek, Nucl.~Phys.~B (Proc. Suppl.) {\bf 37B}, 197 (1994).

\bibitem{JK} M.~Je\.zabek and J.~K\"uhn, \NPB 320 20 1989 .

\bibitem{CLT} D.~Chang, S.-C.~Lee, and P.~Turcotte, hep-ph/9508357.

\bibitem{MP} G.~Mahlon and S.~Parke, hep-ph/9512264.

\end{thebibliography}
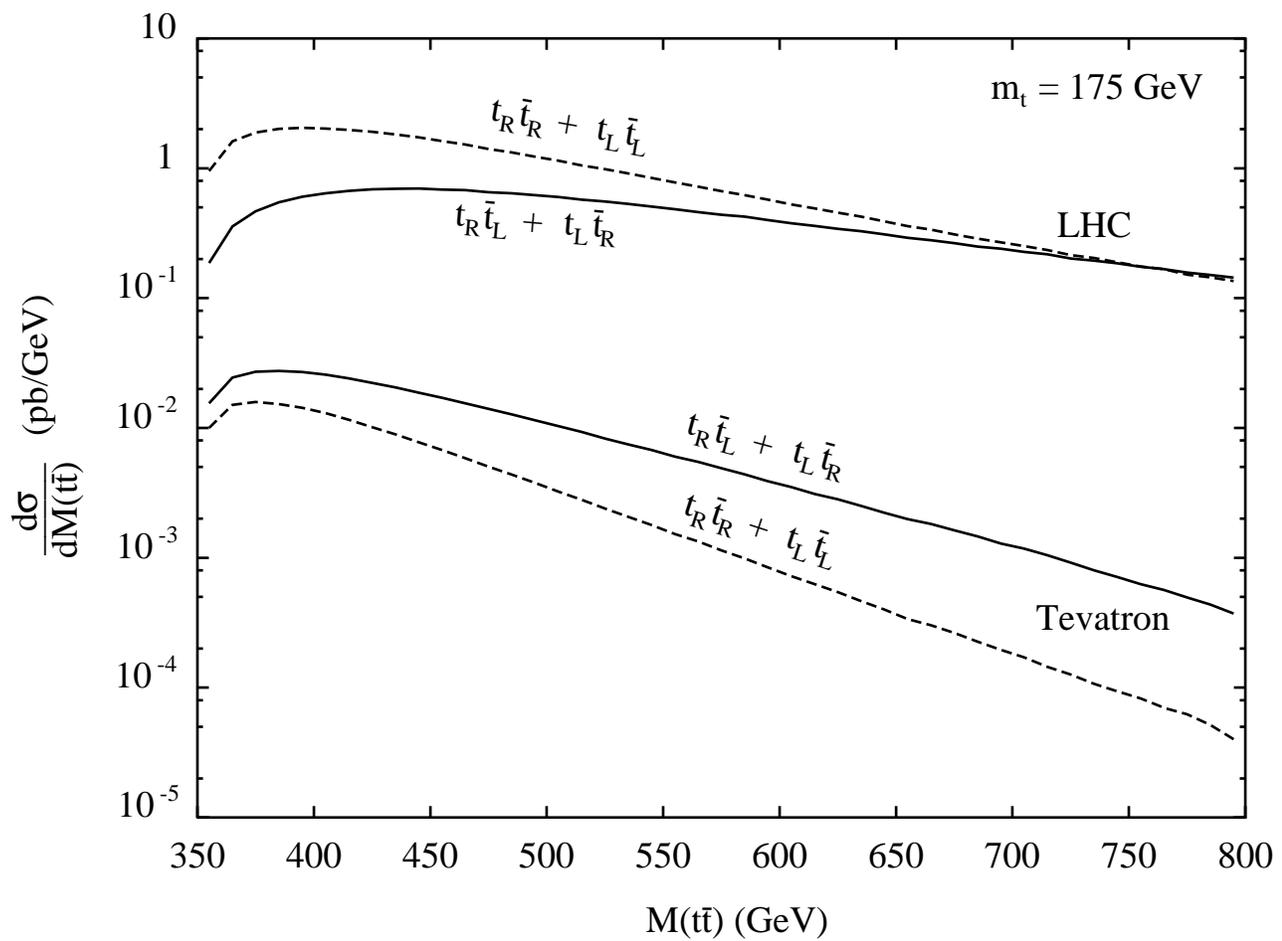
\end{document}